\documentclass[twocolumn]{aastex631}
\usepackage{amsmath}

\begin{document}

\title{A XRISM Observation of the Archetypal Radio-Mode Feedback System Hydra-A: Measurements of Atmospheric Motion and Constraints on Turbulent Dissipation}

\author[0000-0002-8310-2218]{Tom Rose}
\affiliation{Waterloo Centre for Astrophysics, University of Waterloo, Ontario N2L 3G1, Canada}

\author{B. R. McNamara}
\affiliation{Waterloo Centre for Astrophysics, University of Waterloo, Ontario N2L 3G1, Canada}

\author{Julian Meunier}
\affiliation{Waterloo Centre for Astrophysics, University of Waterloo, Ontario N2L 3G1, Canada}

\author{A. C. Fabian}
\affiliation{Institute of Astronomy, Cambridge University, Madingley Rd., Cambridge, CB3 0HA, UK}

\author[0000-0001-5208-649X]{Helen Russell}
\affiliation{School of Physics \& Astronomy, University of Nottingham, Nottingham, NG7 2RD, UK}

\author[0000-0003-0297-4493]{Paul Nulsen}
\affiliation{Center for Astrophysics $|$ Harvard \& Smithsonian, 60 Garden Street, Cambridge, MA 02138, USA}
\affiliation{ICRAR, University of Western Australia, 35 Stirling Hwy, Crawley, WA 6009, Australia}

\author{Neo Dizdar}
\affiliation{Waterloo Centre for Astrophysics, University of Waterloo, Ontario N2L 3G1, Canada}

\author{Timothy M. Heckman}
\affiliation{The William H. Miller III Department of Physics and Astronomy, The Johns Hopkins University,
Baltimore, MD 21218, USA}

\author[0000-0001-5226-8349]{Michael McDonald}
\affiliation{MIT Kavli Institute for Astrophysics and Space Research, Massachusetts Institute of Technology, Cambridge, MA 02139, USA}

\author{Maxim Markevitch}
\affiliation{NASA Goddard Space Flight Center, Code 662, Greenbelt, MD 20771, USA}

\author[0000-0001-9225-6481]{Frits Paerels}
\affiliation{Columbia Astrophysics Laboratory, Columbia University, 538 W. 120th St., New York, NY, 10027, USA}

\author[0000-0002-9714-3862]{Aurora Simionescu}
\affiliation{SRON Space Research Organisation Netherlands, Niels Bohrweg 4, 2333 CA Leiden, The Netherlands}

\author[0000-0003-0392-0120]{Norbert Werner}
\affiliation{Department of Theoretical Physics and Astrophysics, Faculty of Science, Masaryk University, Kotl{\'a}{\v{r}}sk{\'a}, Brno, 61137, Czech Republic}

\author{Alison L. Coil}
\affiliation{Department of Astronomy and Astrophysics, University of California, 9500 Gilman Dr., La Jolla, CA 92093}

\author[0000-0002-2397-206X]{Edmund Hodges-Kluck}
\affiliation{University of Michigan NASA Goddard Space Flight Center: Greenbelt, MD}

\author[0000-0002-3031-2326]{Eric D. Miller}
\affiliation{Kavli Institute for Astrophysics and Space Research, Massachusetts Institute of Technology, MA 02139, USA}

\author[0000-0002-6470-2285]{Michael Wise}
\affiliation{SRON, Netherlands Institute for Space Research, Niels Bohrweg 4, 2333 CA Leiden, The Netherlands}



\begin{abstract}
We present XRISM Resolve observations centered on Hydra-A, a redshift z = 0.054 brightest cluster galaxy which hosts one of the largest and most powerful FR-I radio sources in the nearby Universe. We examine the effects of its high jet power on the velocity structure of the cluster's hot atmosphere. Hydra-A's central radio jets have inflated X-ray cavities with energies upward of $10^{61}~\rm erg$. They reach altitudes of 225\,kpc from the cluster center, well beyond the atmosphere's central cooling region. Resolve's $3\times3$\,arcmin field-of-view covers $190\times190$\,kpc, which encompasses most of the cooling volume. We find a one dimensional atmospheric velocity dispersion across the volume of $164\pm10\,\rm{km\,s}^{-1}$. The fraction in isotropic turbulence or unresolved bulk velocity is unknown. Assuming pure isotropic turbulence, the turbulent kinetic energy is $2.5\,\%$ of the thermal energy radiated away over the cooling timescale, implying that kinetic energy must be supplied continually to offset cooling. While Hydra-A's radio jets are powerful enough to supply kinetic energy to the atmosphere at the observed level, turbulent dissipation alone would struggle to offset cooling throughout the cooling volume. The central galaxy's radial velocity is similar to the atmospheric velocity, with an offset of $-37 \pm 23\,$\,km\,s$^{-1}$.

\end{abstract}



\section{Introduction} 
\label{sec:intro}

\begin{figure*}[]
    \centering
    \includegraphics[width=\textwidth]{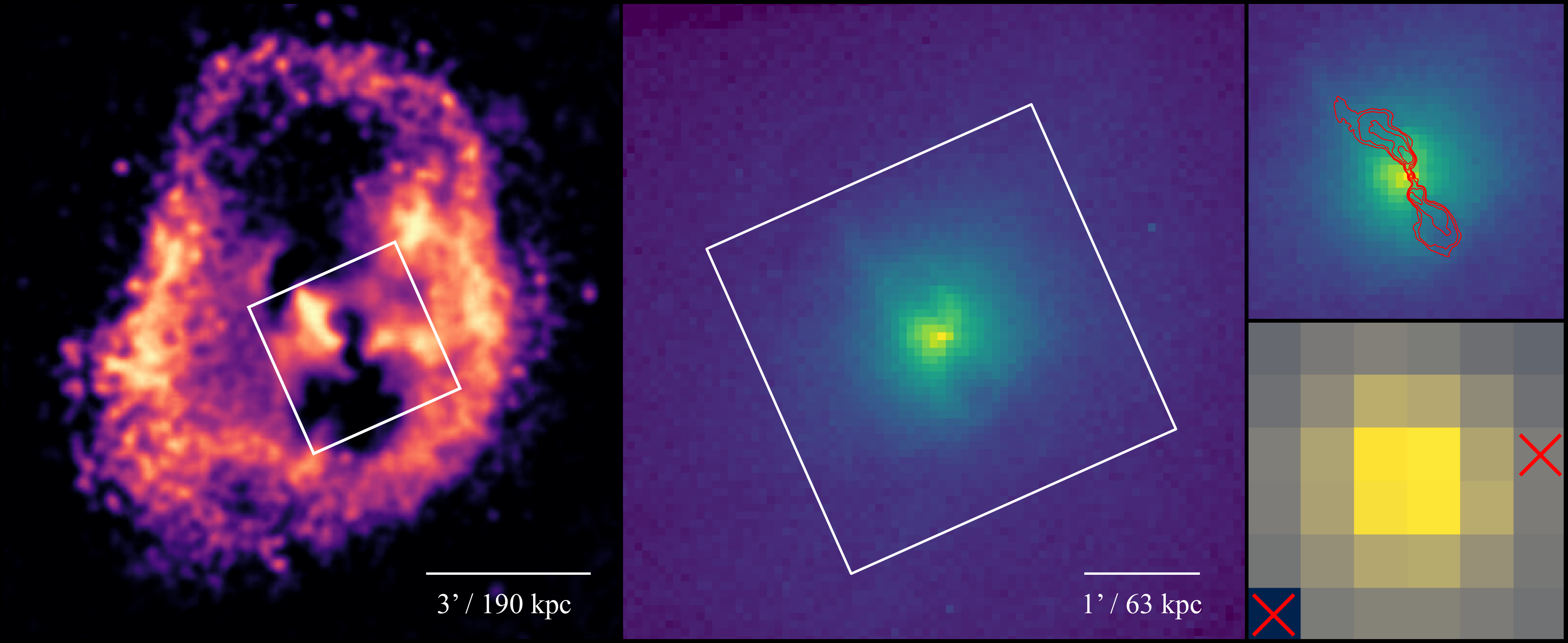}
    \caption{\textbf{Left}: Chandra X-ray residual map centered on Hydra-A after subtraction of a beta model fit to the cluster surface brightness profile, from \citet{Wise2007}. Multiple cavities can be seen out to 225\,kpc. White rectangles show the field-of-view of the XRISM Resolve observation. \textbf{Middle}: Chandra X-ray image centered on the cluster at $0.5-7.0$\,keV. We use `sqrt' stretching to highlight the innermost cavities, formed due to the action of AGN's radio jets and lobes. \textbf{Upper right:} A repeat of the middle image, with radio lobes observed by the JVLA at 5\,GHz (Project 13B-088) shown in red. \textbf{Lower Right}: The $3 \times 3$\,arcmin XRISM observation of Hydra-A, with the standard orientation of the XRISM Resolve footprint. Red crosses indicate the calibration pixel and pixel 27, which is excluded due to  gain jumps.}
    \label{fig:Hydra_A_chandra_and_XRISM_images}
\end{figure*}

Massive galaxies and their host clusters are surrounded by hot atmospheres of low density gas, many of which emit brightly at X-ray wavelengths \citep[e.g.][]{Sarazin1998, Allen2001, Fabian2011}. The power radiated away by this high energy emission can imply cooling rates as high as several hundreds of solar masses per year. Without continual energy input, the hot atmospheres of these clusters would cool and fuel star formation at levels far exceeding those observed \citep[][]{McNamara2012, McDonald2018,DonahueVoit2022}. 

A continual source of heating in galaxy clusters is provided by active galactic nuclei (AGN). AGN efficiently convert accreted matter into feedback power in the form of radio jets and lobes, which distribute their energy throughout the intracluster medium \citep[][]{Voit06,Davis2011}. This feedback power is energetically able to replenish radiative cooling losses and buttress the atmosphere against catastrophic cooling. 

How AGN dissipate their energy is unknown, although several mechanisms have been proposed. They include sound waves \citep[e.g.][]{Sanders2008, Fabian2017,Bambic19}, turbulence in the ICM \citep[e.g.][]{Zhuraleva2014, Mohapatra2019}, cosmic rays \citep[e.g.][]{Fujita2013, Ruszkowski2023}, shock waves \citep[e.g.][]{McCarthy2007, Nulsen05}, the uplift of gas from inner regions \citep[e.g.][]{Zhang2022, Husko2023}, thermal conduction \citep[e.g.][]{Voigt2004}, and heating via the mixing of hot gas from jet inflated X-ray cavities \citep[e.g.][]{Gulkis2012,Hillel2016}.

Among these, turbulent energy dissipation is often favored as the dominant means by which jet energy is spread throughout the intracluster medium, at least in part because it naturally produces a self regulated fueling and feedback loop. If turbulent heating does dominate, velocity widths in the X-ray emitting atmospheres of several hundreds of km\,s$^{-1}$ are expected \citep[][]{Mohapatra2019}. However, some simulations such as those by \citet[][]{Bourne2017} indicate that turbulent heating may be ineffective at heating cluster atmospheres, in which case lower levels of turbulence may be expected.

Measurements of atmospheric turbulence are key to determining the significance of turbulent energy dissipation. Until recently however, only limits on atmospheric velocity widths could be detected, achieved using grating spectrographs \citep[]{Sanders13}. The Hitomi X-ray telescope's micro-calorimeter was launched in 2016 and provided high resolution X-ray spectroscopy. It found velocity broadening in the Perseus cluster atmosphere, in the vicinity of its X-ray cavities, of less than 200\,km\,s$^{-1}$ \citep[][]{Hitomi2016,Hitomi18}, indicating that AGN feedback is able to generate subsonic turbulence contributing roughly 4\% of the total atmospheric pressure. 

The XRISM X-ray Observatory's Resolve microcalorimeter has observed several galaxy clusters with a broad range of AGN jet power, including recently published results for the Abell 2029, Centaurus, Coma and Ophiuchus clusters \citep[][]{XRISM_2025_A2029, XRISM_Centaurus_2025, XRISM2025Coma, XRISM_Ophiuchus2025}. Abell 2029 has a powerful central radio source but no X-ray cavities coincident with its jets and lobes \citep{XRISM_2025_A2029}, while Centaurus contains several cavities formed over a broad epoch of AGN feedback \citep{XRISM_Centaurus_2025}. Coma is a merging cluster without significant AGN jets or X-ray cavities \citep{XRISM2025Coma}. Ophiuchus contains multiple cold fronts and there is evidence of dynamical disturbances in the cluster core \citep{XRISM_Ophiuchus2025}. Despite these differences, XRISM has found relatively low central velocity dispersions in all four ($<210$\,km\,s$^{-1}$), suggesting radio jets and lobes have a tangible yet limited impact on the overall level of turbulence.

Following these initial results, we present XRISM Resolve observations centered on the cluster hosting the archetypal radio-mode feedback galaxy Hydra-A, shown in Figure \ref{fig:Hydra_A_chandra_and_XRISM_images}. Hydra-A is among the largest and most powerful FR1 radio sources known \citep[][]{Lane2004}, providing a test of the effects of high jet power on the velocity structure of the cluster's hot atmosphere. 

Hydra-A’s radio jets have formed three pairs of X-ray cavities \citep[][]{Wise2007}. The youngest and innermost cavities, which are entirely encompassed within our central XRISM pointing, extend 25\,kpc to the north and south of the galaxy’s nucleus. The other, older bubbles are much larger in size and much further from the cluster center. Two are found 100 and 225\,kpc to the north of the radio core. Two more lie 60 and 100\,kpc to the south. This series of bubbles and their associated shock fronts have released energies upward of $\sim 10^{61}$\,erg over the past $200-500$\,Myr, corresponding to a mean power of $\approx 2\times 10^{45}$\,erg\,s$^{-1}$ \citep{Nulsen2005}. This jet power exceeds that of Perseus by more than a factor of 10 \citep{Wise2007, Birzan2004, Rafferty2006, Dunn2006}. 

The cluster's central galaxy hosts a 5\,kpc wide rotating disk of molecular gas, with a smooth velocity gradient and a velocity range of roughly 700\,km\,s$^{-1}$  \citep[][]{Rose2019a, Rose2020}. This molecular disk may be providing fuel for the radio jets \citep[][]{Rose2024_abs}. 

Hydra-A was observed with XRISM to investigate the properties of its hot atmosphere. Using XRISM's 5 eV spectral resolution at 6 keV \citep[][]{Eckart2024}, we place strong constraints on the extent to which the powerful active galactic nucleus has affected the surrounding hot atmosphere. 

Throughout this paper, we assume a flat $\Lambda$CDM Universe with H$_{0}=70$ km s$^{-1}$ Mpc$^{-1}$, \mbox{$\Omega_{M}$=0.3} and \mbox{$\Omega_{\Lambda}$=0.7}. At Hydra-A’s redshift, $z = 0.0543$, there is a spatial scale of 1.056\,kpc/arcsec.

\begin{figure*}
    \centering
    \includegraphics[width=\textwidth]{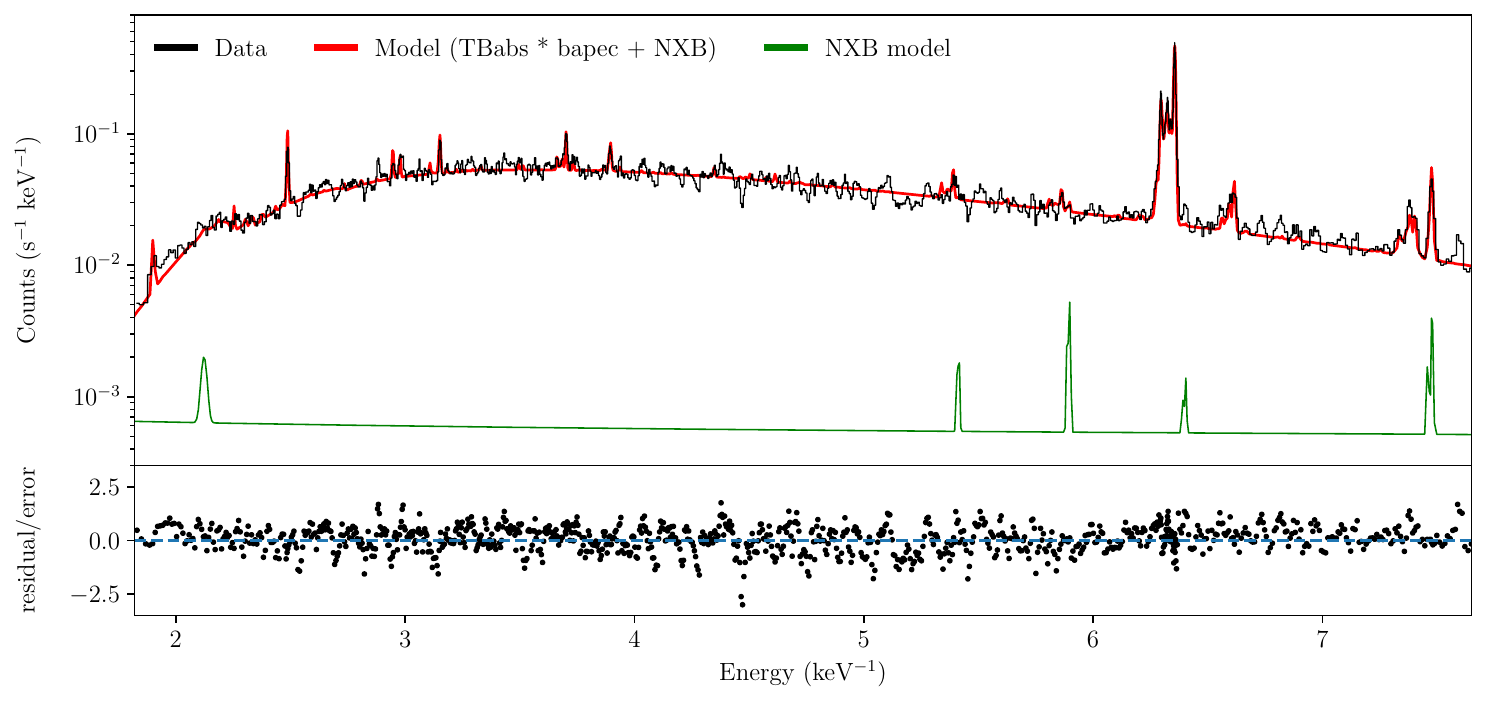}
    \caption{The $1.8-8.0$\,keV spectrum extracted from the XRISM observation centered on Hydra-A and the associated best fit model found over the same energy range. The best fit model contains three components. The (TBabs*bapec + NXB) components are shown in red and the NXB component alone is shown in green. In this figure, the spectrum is grouped to a minimum of 15 counts per bin. However, the best fit model is calculated without re-binning. The soft excess below 2\,keV is likely due to effective area uncertainties as a result of XRISM's closed gate valve.}
    \label{fig:Full_spectrum_and_broadband_fits}
\end{figure*}

\begin{figure}
    \centering
    \includegraphics[width=\columnwidth]{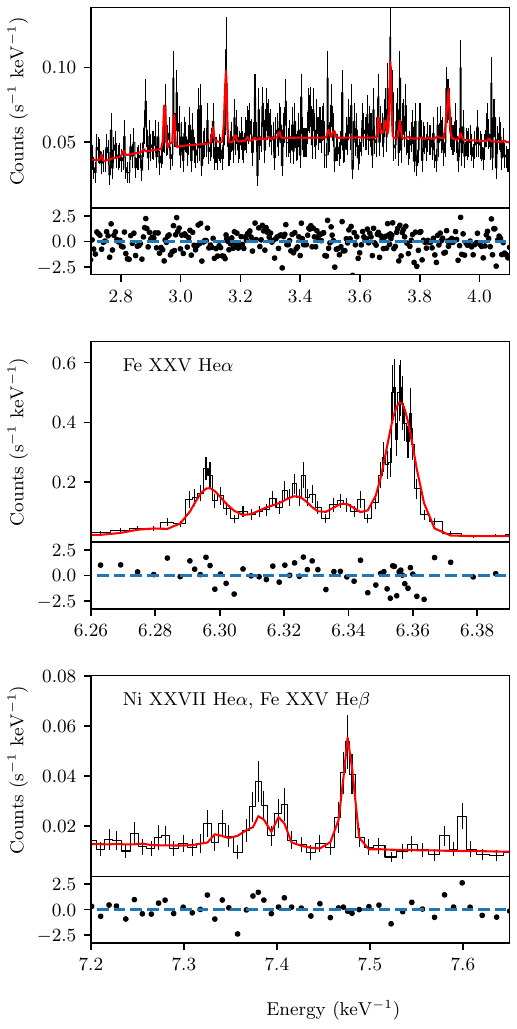}
    \caption{Zoom-ins for the XRISM spectrum of Hydra-A. The emission lines shown in each plot are those used to calculate narrowband fits in Section \ref{sec:narrowband_fits}. Red lines show the broadband fit calculated from $1.8 - 8.0$\,keV. Residuals show the (data - model) / error.}
    \label{fig:individual_line_fits}
\end{figure}
\section{Analysis} \label{sec:intro}
\label{sec:analysis}

\subsection{Observation}

Hydra-A was observed with XRISM in a single observation starting November 27, 2024 (Obs ID: 201070010). The observation was centered at RA = 139.5226 deg, Dec = -12.0951 deg and has a nominal aspect point roll of 114.49 degrees. The Resolve data were processed with \texttt{HEASoft} version 6.34. After screening \citep[as described by][]{XRISM_2024_N132D}, the observation has a cleaned exposure time of 116.3\,ksec.

XRISM velocities in this paper have a barycentric correction of 27\,km\,s$^{-1}$ applied, calculated at JD 2460639 using the equations of \citet{Wright2014}. Since the observation is completed over several orbits, it will also contain low Earth orbit broadening of around 8\,km\,s$^{-1}$. However, assuming this adds in quadrature with the true velocity broadening, this produces only a 0.2\,km\,s$^{-1}$ increase in the observed value.

\subsection{Data extraction}

A spectrum was extracted from the full array after excluding pixel 27, which has shown gain jumps missed by the $^{55}$Fe fiducial cadence. This approach has been followed by other analyses of XRISM data \citep[e.g.][]{XRISM_2025_A2029}. Integrated over the entire spectrum and field-of-view, we detect a photon count rate of $0.230\pm0.001$\,s$^{-1}$ over the 116.3\,ksec exposure time. A review of the Resolve Gain Recovery Report for this observation indicated no issues with the energy scale reconstruction for the remaining pixels.

In our analysis, we include only high-resolution primary events, which account for 94\% of the 2–10 keV events. This fraction ignores low-resolution secondaries, the large majority of which arise from instrumental effects at the low count rates. 

Figure \ref{fig:Full_spectrum_and_broadband_fits} shows the extracted spectrum. XRISM's native resolution is 5\,eV per bin, but for ease of viewing the spectrum in Figure \ref{fig:Full_spectrum_and_broadband_fits} has been binned to 15 counts per bin. Zoomed-in spectra for key emission lines are also shown in Figure \ref{fig:individual_line_fits}.

\subsection{RMF and ARF generation}

A redistribution matrix file (RMF) was generated using the Build 8 \texttt{ftools}, the final pre-launch suite of XRISM software provided to the science team. To make the RMF, we used the CalDB file \texttt{rmfparam\_20190101v006}, which at the time of analysis was the most up to date version available. 

The ancillary response file (ARF) was generated with \texttt{xaarfgen}. The $0.5-7.0$\,keV Chandra image shown in Figure \ref{fig:Hydra_A_chandra_and_XRISM_images} was used as the source model and we produced the ARF with $1\times10^{6}$ photons and a $3'\times3'$ Chandra image whose field-of-view matched our XRISM observation. We note that the energy range of the Chandra image does not exactly match that of our XRISM data. However, as we will come to show, we detect no variation in the distribution of flux from $0.5 - 7$\,keV and we do not expect a significant difference marginally outside this range. The Chandra datasets we use, obtained by the Chandra X-ray Observatory, are contained in the Chandra Data Collection \href{https://doi.org/10.25574/cdc.409}{doi:10.25574/cdc.409}.

XRISM has a point spread function which mixes photons from different spatial origins (see the XRISM proposers' observatory guide for a detailed description). If there is bulk motion within Hydra-A, this could lead to an overestimation of the turbulent velocity in our analysis. Since we work with a single image, we do not account for spatial-spectral mixing from the instrumental point spread function. However, from ray-tracing simulations with \texttt{xrtraytrace}, we estimate that 11\% of the X-ray emission detected in the XRISM pointing originates from outside Resolve's field-of-view. Based on the Chandra observation centered on Hydra-A, we estimate that 72\% of the entire cluster's X-ray flux lies within the XRISM Resolve field-of-view, with the rest being more extended.

We also performed some basic analysis using an ARF generated using a point source model. In the best fit models described in the following subsections, this produces some differences in the normalizations and derived fluxes, but other fitting parameters are consistent within errors.

\begin{table*}
    \centering
    \begin{tabular}{lcccc}
        \hline
         & \multicolumn{4}{c}{Energy Range (keV)} \\
         Parameter & 1.8 -- 8.0 & 2.7 -- 4.1 & 6.15 -- 6.5 & 7.2 -- 7.65 \\
         \hline
         nH (cm$^{-2}$) & $4.04\times10^{20}$ & $4.04\times10^{20}$ & $4.04\times10^{20}$ & $4.04\times10^{20}$ \\
         kT (keV) & $3.6^{+0.1}_{-0.1}$ & $3.5^{+0.2}_{-0.3}$ & $3.34^{+0.21}_{-0.17}$ & $2.8^{+0.5}_{-0.4}$ \\
         Z (Z$_{\odot}$) & $0.45^{+0.01}_{-0.01}$ & $0.42^{+0.08}_{-0.08}$ & $0.39^{+0.03}_{-0.03}$ & $0.43^{+0.08}_{-0.07}$ \\
         Redshift & $0.05421^{+0.00005}_{-0.00002}$ & $0.05410^{+0.00017}_{-0.00014}$ & $0.05421^{+0.00005}_{-0.00003}$ & $0.05434^{+0.00012}_{-0.00014}$ \\
         $\sigma_{v}$ (km\,s$^{-1}$) & $164^{+10}_{-10}$ & $146^{+86}_{-92}$ & $160^{+11}_{-10}$ & $154^{+42}_{-35}$ \\
         N\textsubscript{bapec} & $0.031^{+0.001}_{-0.001}$ & $0.032^{+0.003}_{-0.002}$ & $0.038^{+0.004}_{-0.004}$ & $0.053^{+0.025}_{-0.015}$ \\
         C-stat/dof & 13833/12393 & 3026/2795 & 803/694 & 1036/893 \\
         \hline
    \end{tabular}
    \caption{Fitting parameters for the emission extracted from the XRISM observation centered on Hydra-A, the spectra of which are shown in Figures \ref{fig:Full_spectrum_and_broadband_fits} and \ref{fig:individual_line_fits}. We fit using a (TBabs*bapec + NXB) model for one broad energy range and three narrow energy ranges. For the TBabs component, we freeze the Galactic line of sight hydrogen column density to nH$ = 4.04\times10^{20}\textnormal{cm}^{-2}$ \citep{HI4PI2016}.}
    \label{tab:fits}
\end{table*}

\begin{figure*}
    \centering
    \includegraphics[width=\textwidth]{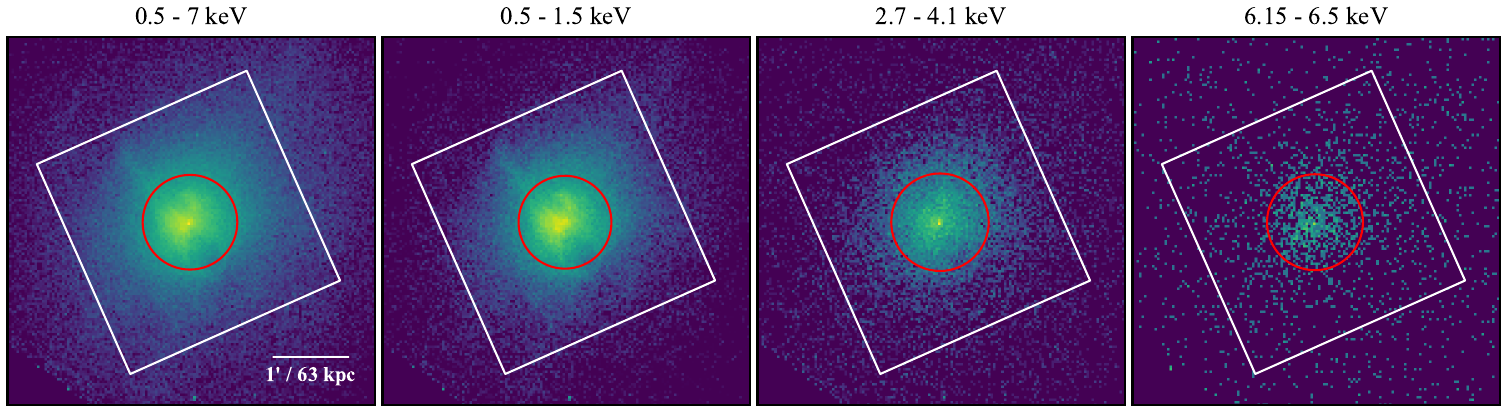}
    \caption{Chandra images of four different energy windows. The left image shows a broad energy range, while the other three are made with narrower energy ranges. Red circles show the radius containing 50\% as much flux as a circular region of 1.5 arcmin radius around the cluster center. From left to right, these radii are 37.1, 36.4, 38.2 and 37.6 arcsec. Therefore, the degree to which the flux is centrally concentrated is fairly constant. All images have the same spatial scale and `sqrt' stretching. White rectangles show the field-of-view of the XRISM Resolve observation.}
    \label{fig:individual_line_images}
\end{figure*}

\begin{figure}
    \centering
    \includegraphics[width=\columnwidth]{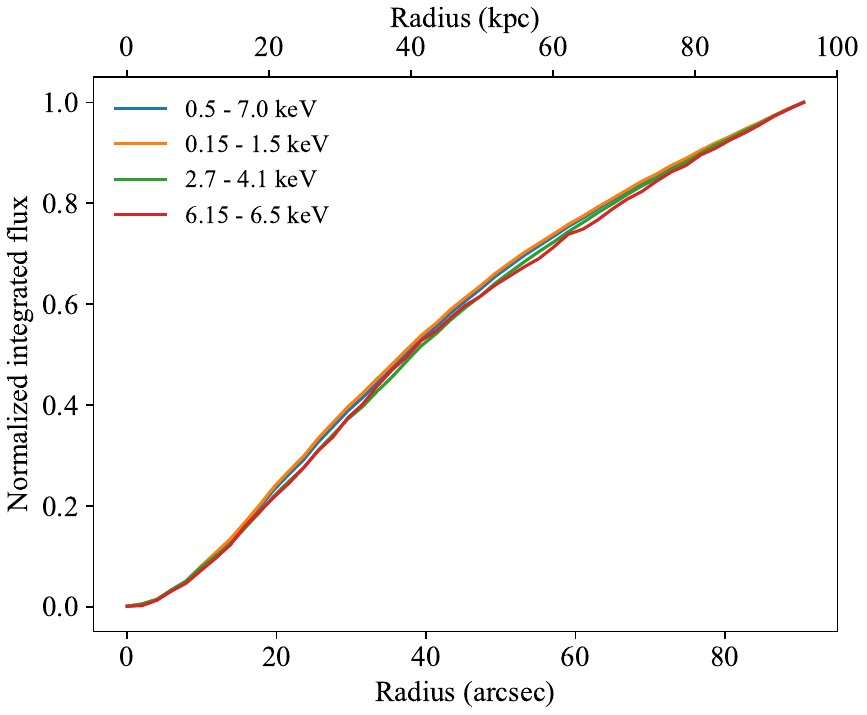}
    \caption{The integrated flux as a function of radius for the four Chandra images shown in Figure \ref{fig:individual_line_images}. The integrated fluxes are normalized to the value at 90\,arcsec, the radius matching the field-of-view of the XRISM image. The flux distribution shows very little variation across the different energy ranges, so the spatial distribution of the flux contributing to the various spectral lines does not change.}
    \label{fig:Chandra_integrated_fluxes}
\end{figure}

\subsection{Broadband fit}

A best fit model was generated for the unbinned spectrum over the range $1.8-8.0$\,keV. These fits can be found in Table \ref{tab:fits}. The key best fit parameters are the temperature of $3.6\pm 0.1$\,keV and line of sight atmospheric velocity width of $164\pm10$\,km\,s$^{-1}$. This velocity broadening is about 17\% of the $\simeq 950~\rm km~s^{-1}$ sound speed. We also find a redshift of $0.05421\pm0.00006$, equivalent to an optical velocity of $16252\pm19$\,km\,s$^{-1}$. 
The uncertainty here is due to the statistical uncertainty of the fit, combined with the 0.3\,eV (15\,km\,s$^{-1}$) systematic uncertainty in the Resolve energy scale at 6\,keV. These two errors add in quadrature. There is also a systematic uncertainty in the velocity dispersion of 3\,km\,s$^{-1}$. This has also been accounted for in the reported uncertainties.

The spectra were modeled in \texttt{XSPEC} with a three component (TBabs*bapec + NXB) model \citep[][]{Arnaud1996}. The emissivities of different elements used in XSPEC are calculated in AtomDB \citep{Foster2012}.

In our three component model, TBabs represents the photoelectric absorption in the gas lying along our line of sight in The Galaxy. $\textnormal{N}_{\textnormal{H}_{2}}=4.04\times10^{20}\,\textnormal{cm}^{-2}$ was taken from the HEASARC online database \citep[][]{HI4PI2016}. The bapec component is a velocity-broadened collisional-equilibrium model of the intracluster medium. We also tested a power law component to represent non-thermal emission associated with the central X-ray AGN. However, it changed the fit negligibly and was excluded from the final model.

Using \texttt{rslnxbgen}, the non-X-ray background (NXB) spectrum was generated from a database of night-Earth data collected by Resolve. This spectrum was weighted based on the distribution of geomagnetic cut-off rigidity sampled during each observation. The strength of the NXB spectrum is at least an order of magnitude less than the source flux and is at its highest relative strength at the extreme energies of the spectrum.

We also tested a two temperature model (i.e. two bapec components). However, this test resulted in 1$\sigma$ errors which were comparable in magnitude to the fitting parameters themselves and did not significantly improve the C-stat/dof.

\subsection{Narrowband fits}
\label{sec:narrowband_fits}

The broadband fit described above, which uses fixed atomic ratios, does not result in accurate fits to all of the spectral lines (e.g. see the bottom panel of Figure \ref{fig:individual_line_fits}). We therefore also performed a series of narrowband fits using the same model, but more restricted energy ranges of $2.7 - 4.1$, $6.15 - 6.5$ and $7.2 - 7.65$\,keV. The resulting best fit models are shown in Table \ref{tab:fits}.

Of particular interest is the FeXXV He$\alpha$ complex, which is the strongest in the spectrum. Judging from the middle panel of Figure \ref{fig:individual_line_fits}, the model from $1.8 - 8.0$\,keV is not an ideal fit to this line complex, with the amplitudes of the y and z lines (between 6.29 and 6.33\,keV) appearing underestimated. We therefore tested narrowband fits which ignored the strongest w line, but found that this did not change the velocity center, change the velocity width, or improve the overall quality of the fits, as judged by the C-stat/dof.

\section{Discussion} \label{sec:intro}

\subsection{Spatial distribution of flux at different energies}

XRISM is sensitive to a range of energies over which several emission lines are detected. These may be formed in different spatial regions. Figure \ref{fig:individual_line_images} shows a broadband image and three narrowband images, from which the spatial distribution of flux at different energies can be inferred. In Figure \ref{fig:Chandra_integrated_fluxes}, we also show the integrated flux as a function of distance from the central X-ray point source. These figures show no appreciable differences in the distribution of flux at different energies. 

Most of the weight in our velocity measurements emerges from the iron features. Figures \ref{fig:individual_line_images} and \ref{fig:Chandra_integrated_fluxes} show no excess iron emission emerging from the bright cavity rims that may have significant expansion velocities. Since half of the iron photons are produced in the bright central region near the cavity rims, the rims do not dominate the signal.

\subsection{Atmospheric velocity structure}

\begin{table}
    \centering
    \begin{tabular}{lccc}
        \hline
         & $z$ & c$z$ & Obs ID \\
        \hline
        MUSE & $0.05434\pm0.00005$ & $16294\pm15$ & 094.A-0859 \\
        ALMA & $0.05431\pm0.00005$ & $16284\pm15$ & 2016.1.01214 \\
        XRISM & $0.05421\pm0.00006$ & $16252\pm19$ & 201070010 \\
        \hline
    \end{tabular}

    \caption{Redshifts and their equivalent optical velocities in km\,s$^{-1}$, measured by: 1) MUSE, using stellar absorption lines seen at the brightest cluster galaxy center \citep[][]{Rose2019a}, 2) ALMA, using CO(2-1) emission from the molecular disk \citep[][]{Rose2019a} and 3) XRISM, found in this paper using the entire central pointing at 1.8-8.0\,keV and after a barycentric correction of +27\,km\,s$^{-1}$.}
    \label{tab:redshifts}
\end{table}

Three accurate redshifts for Hydra-A are presented in Table \ref{tab:redshifts}. A MUSE redshift is found using stellar absorption lines at the core of the brightest cluster galaxy \citep[][]{Rose2019a}. The ALMA redshift is for the center of the molecular gas disk \citep[][]{Rose2019a}.

The MUSE and ALMA redshifts, which respectively trace stars and molecular gas, are less than $1\sigma$ apart. However, the hot gas traced by XRISM has a slightly lower redshift. 

The bulk velocity is defined as

\begin{equation}
    v_{\rm{bulk}} = \frac{c(z - z_{\rm{BCG}})}{1 + z_{\rm{BCG}}}.
\end{equation}

Here, $z$ is the redshift of the hot atmosphere, measured here with XRISM to be $0.05421\pm0.00006$. $z_{\rm{BCG}}$ is the redshift of the brightest cluster galaxy, found to be $0.05434\pm0.00005$ using the MUSE observations. These values give a bulk velocity of $-37\pm23$\,km\,s$^{-1}$. 

Relative motion between a central galaxy and its cluster's hot atmosphere changes the effective freefall time and $\rm{t}_{\rm{cool}} / \rm{t}_{\rm{freefall}}$ ratio of cooling gas, which is the basis for precipitation models of thermally unstable cooling \citep[e.g.][]{Voit2015}. A velocity difference which effectively increases the gas freefall time would be similar to lifting \citep[e.g.][]{McNamara2016,DonahueVoit2022}, thus accelerating thermally unstable cooling. However, effects of this kind in Hydra-A would be subtle due to its modest bulk velocity offset.

\subsection{Turbulent Dissipation of Jet Energy}
Turbulent dissipation of jet power could offset cooling if the atmosphere's turbulent kinetic energy density is sufficiently high and if the turbulent speeds and injection scales permit the energy to cascade down to small scales and dissipate throughout the cooling volume before it is radiated away \citep{Fabian2017,Bambic18}. 

The measured atmospheric velocity dispersion of $164\pm 10\,\rm{km\,s}^{-1}$ is surprisingly low considering Hydra-A's high jet power, which is among the most powerful in the nearby Universe. While the jet power of Hydra-A's inner two bubbles is comparable to the inner bubbles in Perseus, its total jet power is ten times that of Perseus \citep{Wise2007, Birzan2004, Rafferty2006, Dunn2006}. Nevertheless, their atmospheric velocity dispersions are similar. If turbulence is dissipating jet power, we would naively expect higher jet power to yield higher turbulent velocities. The velocity dispersion may rise with altitude near the powerful outer bubbles, but confirmation must await a future outer pointing.  

The jet energy associated with the inner two bubble pairs is $E_{\rm jet} = 2.4\times 10^{60}\,\rm{erg}$ \citep[][]{Wise2007}. Multiple bubbles with a range of ages indicate continual power input with a frequency of $\sim 10^8\,\rm yr$ \citep{Vantyghem2014}. Hydra-A is therefore powerful enough to continually offset cooling. We examine here whether this power can be dissipated by turbulence and heat the atmosphere before it is radiated away.

The $3\times 3$\,arcmin XRISM footprint encloses a radius of approximately 94\,kpc, which is a substantial fraction of the $\simeq 100 ~\rm kpc$ cooling radius \citep{Birzan2004, Rafferty2006, Dunn2006}. The atmospheric gas mass within the footprint is ${\rm M}_{\rm atm} = 1.5\pm 0.1
\times 10^{12}\,{\rm M}_\odot$. Assuming an isotropic turbulent velocity that matches our measured value of $\sigma \simeq 160$\,km\,s$^{-1}$, the kinetic energy is \begin{align}
E_{\rm atm} &= 3\times 10^{59} \left({M_{\rm atm} \over 10^{12}\,\rm M_\odot}\right) \left({\sigma \over 100\,\rm{km\,s}^{-1}}\right)^2 ~\rm{erg}.\
\end{align}
\noindent
This figure, calculated as ${3 \over 2} M_{\rm atm}(<r) \sigma^2$, gives a kinetic energy of $E_{\rm atm}=1.1\pm 0.2\times 10^{60}~\rm{erg}$. 
The formal error here is due primarily to the uncertainty in $\sigma$.    

The turbulent kinetic energy is therefore 2.5\% of the $L_{\rm x} = 2 \times 10^{44}\,\rm{erg\,s}^{-1}$ thermal energy radiated away within the (94 kpc)$^3$ volume over the $7\times 10^9\,\rm yr$ cooling time at that radius \citep[][]{Rafferty2006}.  The ratio of kinetic to thermal energy density 
${E_k / E_{th}} = \mu m_p \sigma^2 /kT\simeq 0.047$, or 4.7\%, is twice as large, because it does not account for radiative cooling across the volume. This value is  similar to but slightly larger than for the Abell 2029 \citep[][]{XRISM_2025_A2029} and Centaurus clusters \citep{XRISM_Centaurus_2025}. In order to offset cooling, jet power must then be thermalized throughout the volume in roughly $t_r \sim 0.025 \times 7\times 10^9\,\rm yr= 1.7\times 10^8\,\rm yr$.
This timescale is consistent with the replenishment timescale or duty cycle of the radio bubbles in Hydra-A \citep[][]{Vantyghem2014}. The jets are therefore able to replenish the atmospheric kinetic energy observed within the footprint. But the tolerance is tight and issues remain.

For example, turbulent energy transmitted at $v\sim 160\,\rm{km\,s}^{-1}$ would take at least $6\times 10^8\,\rm yr$ to reach the cooling radius, roughly three times longer than the replenishment timescale. This is the shortest possible timescale for turbulence generated at the center of the atmosphere to propagate throughout the cooling volume, and it is likely considerably longer \citep{Fabian2017}. \citet{Nulsen2005} found the density profile in the cooling region roughly scales as $\rm r^{-1}$. The temperature varies slowly, implying the atmospheric cooling time scales almost linearly with radius. Therefore, the minimum propagation speed of turbulence required to offset cooling is roughly constant at $\sim 500 ~\rm km~s^{-1}$, which is much higher than we observe. This implies that turbulence must be generated efficiently throughout the volume by the bubbles to offset cooling.

Turbulence may be generated as the bubbles rise, which would then be thermalized at higher altitudes. The mean bubble rise times for the inner four cavities roughly enclosed by the XRISM pointing are $5 \times 10^7\,\rm yr$ to reach altitudes of 25\,kpc, and $1.2 \times 10^8\,\rm yr$ to $ 2 \times 10^8\,\rm yr$ to reach altitudes of 60 and 100\,kpc, respectively \citep[][]{Wise2007}. The terminal bubble speeds inferred from these rise times are considerably higher than the turbulent velocity. Therefore, the jet energy has time to be transmitted throughout the cooling volume. But the enthalpy released must in turn generate turbulence that is quickly dissipated. An off-nuclear pointing toward the outer, more powerful radio bubbles would be essential to determine whether bubbles continue to generate turbulence as they rise.

The heating rate due to turbulent dissipation with a Kolmogorov spectrum may be expressed as $\dot E\sim \rho v_{\rm turb}^3/l$ where $l$ is the injection scale. In this approximation, energy is injected throughout the cooling volume at the injection scale and cascades to smaller scales at a constant rate over the timescale $\tau_{\rm turb}\simeq l/v_{\rm turb}$. The Kolmogorov spectrum scales as $v \propto l^{1/3}$. Therefore $\dot E$ is constant on spatial scales smaller than $l$.

This approximation is straightforwardly applied to independent velocity measurements made on multiple (many) lines of sight for which a velocity structure function and injection scale can be measured. Examples include X-ray surface brightness fluctuation analyses \citep{Zhuraleva2014} and emission line velocities measured with integral field spectrometers \citep{Li2020, Li2025}. In these instances, the injection scale is estimated at the turnover of the velocity structure function if one or more exist, or by more complex methods in the case of surface brightness analyses. However, the observed slopes of nebular and molecular velocity structure functions are steeper than 1/3 \citep[]{Li2020, Ganguly2024, Li2025}, and slopes estimated by surface brightness fluctuations have large uncertainties \citep{Zhuraleva2014, Li2025}. Some simulations of hot atmospheres show anisotropic gas velocities \citep[][]{Vazza22}, or slopes that deviate significantly from 1/3, resulting in a large uncertainty in the dissipation rate \citep{Mohapatra2022,Fournier2024,Fournier25}. Recently, \citet[][]{Vazza2025} have argued that the Kolmogorov-like turbulence measured in their simulations of a Coma-like cluster give velocity structure functions and line widths compatible with those measured by XRISM.

A single XRISM pointing alone cannot constrain $l$ and $\sigma(l)$ \citep[cf.,][]{ZuHone16}. Assuming an injection scale without velocity measurements on or below that scale risks violating energy conservation. We therefore assume an effective scale (XRISM Collaboration 2025, in prep.) as the length along which 50\% of the light from the source is collected. Based on Figures \ref{fig:individual_line_images} and \ref{fig:Chandra_integrated_fluxes}, this length is $l_{\rm eff}\simeq 78$ kpc. We then express the power injected as a function of effective scale over the XRISM image as:
\begin{equation}
  \dot E \simeq {3 \over 2} {v_{\rm turb}^3 \over l_{\rm eff}} M(r). 
\end{equation}
$M(r)$ is the atmospheric gas mass subtended by the image and $v_{\rm turb}$ = $\sigma$. The turbulent dissipation timescale, which is not well understood in a weakly-magnetized, stratified atmosphere, is assumed to be $l_{\rm eff}/v_{\rm turb}$. 

This expression yields an energy dissipation rate of $7.6\times 10^{43}\,\rm{erg\,s}^{-1}$. The cooling luminosity subtended by the image is $2.7\times 10^{44}\,\rm{erg\,s}^{-1}$, which exceeds the turbulent energy dissipation rate by six times. 

On the other hand, if one assumes an injection scale adjusted downward by a factor of six so that $l_{\rm eff}\simeq 13 ~\rm kpc$, the turbulent power injected on this scale would equal cooling losses. 

Since the recent and previous generation of bubbles nearly fill the footprint (Figure \ref{fig:Hydra_A_chandra_and_XRISM_images}) an injection scale of this size may be possible. If instead turbulence is generated primarily while the cavities are inflating nearer to the nucleus, the propagation timescale and other issues associated with dumping a huge amount of energy into a small volume on a short timescale must be considered. Of course, this argument is conjectural because we cannot measure $\sigma$ and $l_{\rm eff}$ with XRISM on such small scales.

\subsection{Feasibility of turbulent dissipation in Hydra-A}

While the observed levels of turbulent dissipation and cooling are within a factor of a few of each other, the tolerances are uncomfortably tight due to low turbulent speeds chasing rapidly cooling gas. Moreover, any complexity which has been unaccounted for in our analysis, such as projection effects, multi temperature phases, photon leakage and bulk motions, would only lead us to overestimate the atmospheric turbulence and turbulent energy dissipation. This puts further strain on the uncomfortably tight tolerances.

For example, we have assumed $\sigma$ represents ideal isotropic turbulence which is unlikely to be accurate. Much of $\sigma$ may be unresolved bulk motion that may not quickly dissipate. Figures \ref{fig:individual_line_images} and \ref{fig:Chandra_integrated_fluxes} show that much of the light in the footprint is emerging from the vicinity of the central cavities. \citet[][]{Wise2007} estimated the expansion rates of the cavities to be $\sim 350 ~\rm km~s^{-1}$. This figure is close to the FWHM$\simeq 375~\rm km~s^{-1}$ of the emission lines in the central pointing. Therefore, a significant fraction of the motion assigned here to turbulence could be unresolved bulk motion. The kinetic energy released by bulk motion would eventually thermalize. Whether it would do so quickly enough is unclear.

The data indicate that Hydra-A's radio jets are churning up the atmosphere and could continually sustain the kinetic energy in the atmosphere. However, it is unclear that turbulent dissipation is the primary heating mechanism in this system. Other heat sources, in addition to turbulence, may be required to balance cooling across the entire cooling volume. If dissipation of turbulence and bulk kinetic energy are the primary heating mechanisms, the relatively low velocity dispersion found in the vicinity of Hydra-A's radio source and the tight energy tolerances must reflect a tightly-coupled and extraordinarily gentle feedback loop.

\subsection{Final Remarks}

We found that turbulent heating alone, while likely significant on some scales, would struggle to offset radiative cooling in Hydra-A's hot atmosphere. However, the velocity measurement reported here is a snapshot in time. The timescale for turbulence to cascade and dissipate into heat may be of the same order as the timescales required to radiate away the turbulent energy and for the AGN to replenish it. This problem requires averaging over a larger sample before more definitive conclusions can be made. While the sub-dominance of turbulent heating found here is consistent with other studies \citep[e.g.][]{Mohapatra2019,Yang16,Bambic18}, turbulence and bulk motions surely play a critical role in the heating and cooling cycles of cluster atmospheres \citep[][]{Voit18}.  Another factor that may be important is the growing evidence for peculiar motion between the atmosphere, colder gas phases, and the central galaxy found in several systems \citep[e.g.][]{XRISM_Centaurus_2025,Gingras2024}. A fuller picture of how radio-mode feedback heats atmospheres should unfold in the next few years as XRISM observations accumulate. 

\begin{acknowledgments}

T.R. thanks the Waterloo Centre for Astrophysics and generous funding to B.R.M. from the Canadian Space Agency and the National Science and Engineering Research Council of Canada. HRR acknowledges support from an Anne McLaren Fellowship from the University of Nottingham. NW is supported by the GACR grant 21-13491X. EDM acknowledges support from NASA grants 80NSSC20K0737 and 80NSSC24K0678.

We gratefully acknowledge the hard work over many years of all of the engineers and scientists who made the XRISM mission possible.

This research made use of \texttt{Astropy} \citep{the_astropy_collaboration_astropy_2013,the_astropy_collaboration_astropy_2018}, \texttt{Matplotlib} \citep{hunter_matplotlib_2007}, \texttt{numpy} \citep{walt_numpy_2011,harris_array_2020}, \texttt{Python} \citep{van_rossum_python_2009}, \texttt{Scipy} \citep{jones_scipy_2011,virtanen_scipy_2020} and \texttt{Aplpy} \citep[][]{aplpy}. We thank their developers for maintaining them and making them freely available.

\end{acknowledgments}

%

\vspace{5mm}





\newpage 


\bibliographystyle{aasjournal}



\end{document}